\documentclass[aps,prl,reprint,superscriptaddress,showkeys,longbibliography]{revtex4-1}
\usepackage{graphicx,amsmath,color}

\begin{document}

\title{From Fano resonances to bound states in the continuum in dipole arrays at THz frequencies}
%: experimental evidence for Au rod dimer metasurfaces at THz frequencies}
\author{Diego R. Abujetas} 
\email{diego.romero@iem.cfmac.csic.es, N.J.J.Hoof@tue.nl}
\thanks{These two authors contributed equally.}
\affiliation{Instituto de Estructura de la Materia (IEM-CSIC), Consejo Superior de Investigaciones Cient\'{\i}ficas,\\ Serrano 121, 28006 Madrid, Spain}
\author{Niels van Hoof}
%\email{N.J.J.Hoof@tue.nl}
%\thanks{These two authors contributed equally.}
\affiliation{Institute for Photonic Integration, Department of Applied Physics, Eindhoven University of Technology, P.O. Box 513, 5600 MB Eindhoven, The Netherlands}
\author{$\! ^{,*}$ Stan ter Huurne}
\affiliation{Institute for Photonic Integration, Department of Applied Physics, Eindhoven University of Technology, P.O. Box 513, 5600 MB Eindhoven, The Netherlands}
\author{Jaime G\'omez Rivas}  
\affiliation{Institute for Photonic Integration, Department of Applied Physics, Eindhoven University of Technology, P.O. Box 513, 5600 MB Eindhoven, The Netherlands}
\author{Jos\'e A. S\'anchez-Gil}
\email{j.sanchez@csic.es}
\affiliation{Instituto de Estructura de la Materia (IEM-CSIC), Consejo Superior de Investigaciones Cient\'{\i}ficas,\\ Serrano 121, 28006 Madrid, Spain}

\date{\today}

\begin{abstract} 

Fano resonances and bound states in the continuum (BICs) exhibit a rich phenomenology stemming from, respectively, their asymmetric line shapes and infinite quality factors. Here, we show experimentally and theoretically that rod dimer metasurfaces exhibit narrow (high-Q) Fano resonances at THz frequencies. These resonances evolve continuously into a BIC as the rods in each dimer become identical. We demonstrate analytically that this is a universal behavior occurring in arrays of dimers consisting of detuned resonant dipoles. Fano resonances arise as a result of the interference between broad and narrow lattice dipole resonances, with high-Q factors tending to infinity in the detuning parameter space as the narrow lattice resonance becomes a BIC for identical resonant dipoles. Similar configurations can be straightforwardly envisioned throughout the electromagnetic spectrum leading to ultrahigh-Q Fano resonances and BICs of interest in photonics applications such as sensing and lasing.
\end{abstract}

\maketitle

\textit{Introduction.}\textemdash Narrow resonances in optics and electromagnetism are extremely appealing  both from fundamental standpoints (high-Q factors, long-lived states, and energy concentration), and from the obvious implications that the resulting phenomenology may have in photonics  applications (e.g. sensing, lasing, non-linear effects, enhanced spectroscopies, etc.). Much interest has been devoted in the particular case of Fano resonances: asymmetric, typically narrow line shapes that arise as the interference of a narrow dark resonance with a broad bright one \cite{Miroshnichenko2010,LukYanchuk2010,Limonov2017}, the latter being a continuum in the context of atom physics as introduced by Fano. 
%\cite{Fano1935}.
Among the variety of configurations that exhibit Fano resonances \cite{Giannini2011,Rahmani2012,Fan2014,Verellen2014,Abujetas2017}, sub-wavelength arrays of scatterers called metasurfaces are especially interesting for their enhanced collective response, favoring experimental feasibility and applications \cite{LukYanchuk2010,Wu2011,Singh2014a,Yang2015,Yuan2017,Limonov2017}.

Other type of resonant states, not necessarily asymmetric, called bound states in the continuum (BICs), have attracted much interest lately in Physics for their (theoretically) infinite  Q factor. These states are leaky modes that in a certain limit of some parameter space cannot couple to any radiation channel \cite{Hsu2016a}. In order to trap light in such nearly-zero-linewidth electromagnetic modes in photonics, a common approach is to exploit metasurfaces \cite{Marinica2008,Hsu2014,Bulgakov2017,Taghizadeh2017,Doeleman2018,Koshelev2018}, i.e. sub-wavelength arrays (in the non-diffractive region) where only the specular reflection/transmission channels are allowed by symmetry. Such specular channels can be suppressed by tuning the parameters of the system  in various manners, leading to symmetry-protected BICs. In turn, achieving robust BICs and quasi-BICs can be of interest for engineering cavities with arbitrarily high Q-factor for lasing \cite{Kodigala2017,Ha2018}.
%; when coupling is forbidden by the destructive interference between two resonances, the resulting mode is known as a Friedrich–Wintgen BIC \cite{Hsu2016,FriedrichH1985}.

The condition for BICs can be fulfilled by designing an appropriate metasurface that exhibits high quality modes, which can be continuously tuned through a suitable parameter. An elegant system that we propose to study with those properties is a lattice of gold rod resonators. Sub-wavelength metallic rods are shown to have dipolar-like $\lambda/2$-resonances, leading to a variety of resonant phenomena such as surface plasmon lattice resonances in the optical domain \cite{GarciadeAbajo2007,Giannini2010b,Evlyukhin2010a,Rodriguez2011,Meinzer2014,Baur2018,Habib2018} and  electromagnetically-induced transparency at optical and THz frequencies \cite{Chiam2009,Bozhevolnyi2011,Gu2012,Rodriguez2013,Zhu2013,Schaafsma2016,Halpin2017,Halpin2017a}.  

Here, we experimentally demonstrate that metasurfaces consisting of asymmetric, sub-wavelength gold rod dimers, exhibit strong and narrow Fano resonances in the THz transmission spectra. Such Fano resonances become narrower as the dimensions of the dimer rods approach each other, disappearing when rods are identical as a clear signature of BICs. Apart from full numerical calculations in agreement with the experimental results, a simple model for arrays of detuned-resonant-dipole dimers is developed. Hereby, we show analytically that such arrays hybridize both dipolar resonances in the form of broad and narrow lattice resonances, whose interference leads to asymmetric Fano resonances in the zero-order transmittance. This dark Fano resonance is shown in the parameter space of dipole-detuning to get infinitely narrow, becoming an infinite-Q BIC for zero detuning. These calculations fully explain the experimental results in this wider theoretical context of detuned-dipole arrays.

\textit{Experiments.}\textemdash Using optical lithography, metal deposition, and lift-off, we have fabricated samples containing 2D periodic lattices of gold rods on top of a 1.5 mm thick amorphous quartz substrate. These samples were clamped against another 1.5 mm substrate with index matching liquid in between to suppress unwanted reflections, making the total sample thickness 3 mm. Before the thermal evaporation of the 100 nm gold layer, a 3 nm Ti adhesion layer was deposited. All the samples consist  of 2D periodic lattices  of two gold rods per unit cell. The lattice has a square symmetry with a pitch of $a=b=300$ $\mu$m in both directions. One of the rods has a fixed dimension of 200 $\mu$m $\times$ 40 $\mu$m, while the dimension of the other rod is different for each sample that is investigated. The long dimension of the rods is aligned with the $y$-axis, and the rods inside the unit cell are separated by a distance d$_x$ along the $x$-axis. Both rods of the unit cell support $\lambda/2$-resonances in the THz range, given by their length. The resonance frequency is controlled by the long axis of the rods, so we classify each lattice by the length of the second rod, $L_{2}$, covering the range $L_{2} = 125 - 250$ $\mu$m in steps of 25 $\mu$m while keeping $L_{1} = 200$ $\mu$m constant.

\begin{figure}
\includegraphics[width=1\columnwidth]{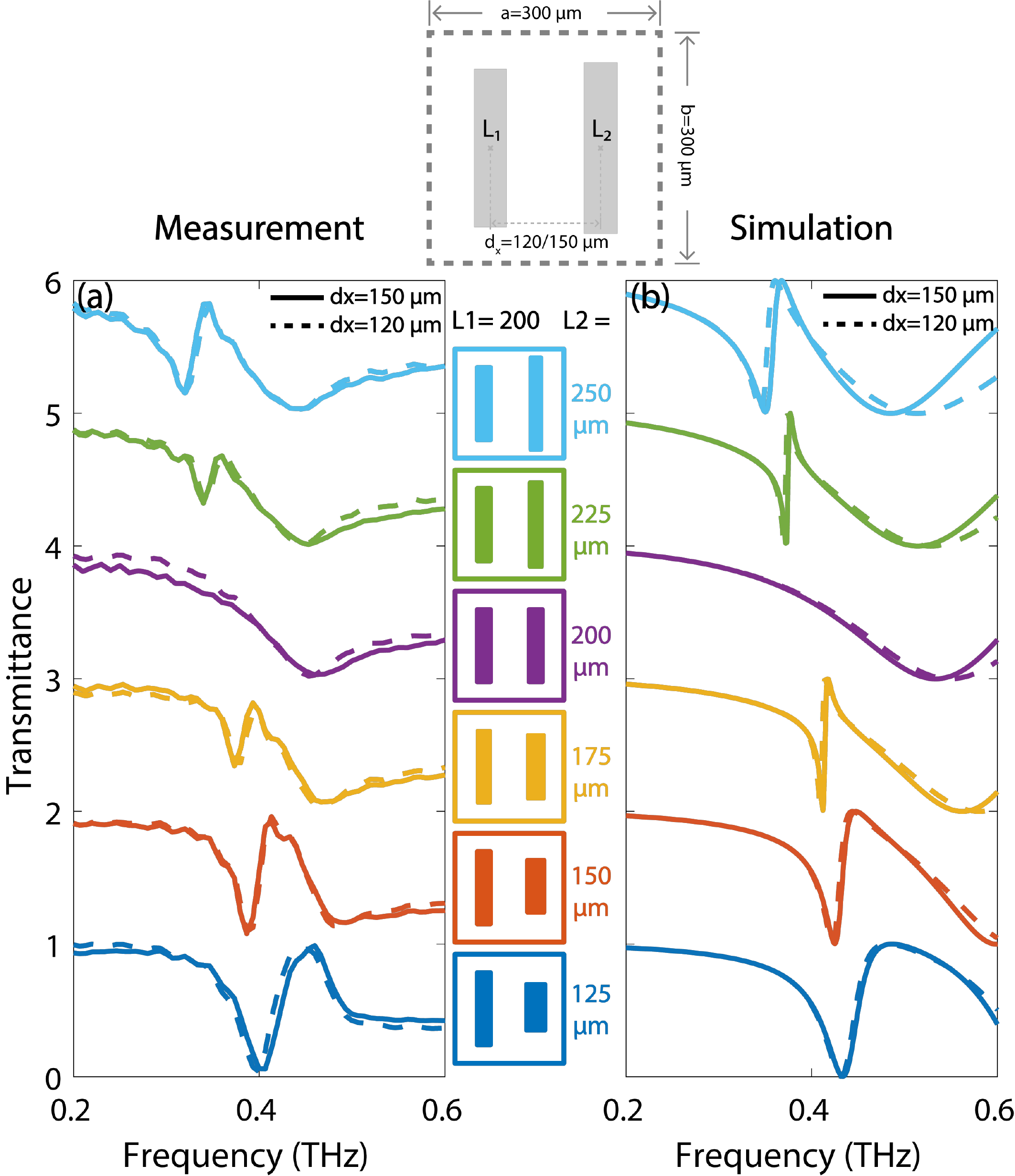}
\caption{(a) Measured transmittance spectra for square lattices  ($a=300$ $\mu$m) of two gold rods per unit cell, deposited on a quartz substrate, with two different rod separations:  $d_x = a/2$ (solid curves) and  $d_x = 2a/5$ (dashed curves). One of the rods has  fixed dimensions of  L$_1=200\,\mu$m and $w_1=40\,\mu$m, while the other  varies as shown in the center insets, with $L_{2}(\mu$m)$ = 125, 150, 175, 200, 225, 250$, while keeping the surface area fixed: $L_2w_2=L_1w_1$. All rod thicknesses are $t=0.1\,\mu$m. (b) Transmittance spectra numerically calculated through SCUFF \cite{SCUFF1,SCUFF2} for the same geometrical parameters, but considering gold rods as planar perfectly conducting rectangles embedded in a uniform medium (see text). Curves are offset by 1 for each different $L_2$.}
  \label{fig:DDexp1}
\end{figure}

Two different sets of samples have been measured, with separation between  rods  fixed at $d_x$ = 120 and 150 $\mu$m. THz transmittance spectra of these samples have been measured at normal incidence using a 4-f far-field THz time domain spectrometer (Menlo TeraK15). The transmittance of the lattices for different $L_{2}$ is shown in Fig.~\ref{fig:DDexp1}(a). Numerical simulations calculated through SCUFF \cite{SCUFF1,SCUFF2} (open-source software package for analysis of electromagnetic scattering problems using the method of moments) are shown in Fig.~\ref{fig:DDexp1}(b), corresponding to  transmittance spectra at normal incidence  for the same geometrical parameters, but considering gold rods as planar perfectly conducting rectangles embedded in a uniform medium with $n=1.55$ (the average of those of air and the supporting quartz substrate \cite{Yu2017a}). There is a shift between the Fano resonances simulated and the measurements, which can be corrected by slightly adjusting this refractive index. Good qualitative, and nearly quantitative, agreement is observed between the measurements and the simulations.

The more relevant feature in Fig.~\ref{fig:DDexp1} is that for $L_{2} \not = 200$ $\mu$m all spectra present a strong Fano resonance with a narrow asymmetric line shape, with a high contrast approaching zero-to-total transmittance. 
The resonance moves toward smaller frequencies as $L_{2}$ is increased, and disappears for $L_{2} = 200$ $\mu$m. In addition, as the rods become more similar in size, the resonance becomes narrower; it should be mentioned, though, that the resonance narrowing is not clearly visible in the measurements for $L=175, 225$ $\mu$m due to the limited frequency resolution in the measurements. The Fano resonance can be understood in terms of the lattice resonances that the array supports: a bright, broad symmetric mode and a dark, narrow antisymmetric mode \cite{Schaafsma2016}. The symmetry of these modes is preserved in the lattice and they interfere either destructively or constructively, producing a characteristic dip/peak pair in the spectra. Moreover, the dark mode becomes inaccessible at $L_{2} = 200$ $\mu$m and the interference disappears; as we will show below, the width of the resonance thus tends to zero, leading to a BIC that cannot be detected in the far field.

However, it should be noted that, when $d_x=150\,\mu$m, the BIC state is connected to the guided mode that the lattice supports. Indeed, for $L_{2} = 200$ $\mu$m both rods are identical, so that the array becomes a single rod lattice with lattice constant halved along the $x$-axis (recall that the lattice pitch is 300 $\mu$m). We call this condition as the "symmetric lattice". The Brillouin zone (and the reciprocal primitive vectors) is doubled when $d_x = a/2$ and the guided mode (below the light line) is linked to the BIC (above the light line). 
%In addition, for both modes (guided mode/BIC) the field at each rod is out of phase (antisymmetric). 
Furthermore, the number of resonances that the lattice supports can be associated to the number of particles per unit cell.  For the identical symmetric lattice, i.e. one particle per unit cell or $L_2=200\,\mu$m and $d_x=150\,\mu$m, there is only a bright (symmetric) mode. Therefore, for the symmetric lattice it is difficult to relate the BIC with a real state.
%rather, it seems to be a manifestation of the guided mode.  

Nonetheless, for a lattice with $d_x=120\,\mu$m, where the symmetric lattice is not recovered at $L_{2}=200\,\mu$m for equal rods, the same details hold. The BIC is no longer directly connected to any guided mode, so that the system has two well defined lattice modes and a true BIC emerges. This is further verified through numerical calculations at oblique incidence (see Supplemental Figure S1), which reveal a Fano resonance emerging for equal rods only in the case of the asymmetric lattice ($d_x=2a/5$); yet another evidence of the true BIC behavior.

Interestingly, if we calculate the scattering efficiencies of the isolated dimers (see Supplemental Figure S2 for three cases), the spectra reveal roughly speaking  two broad lobes corresponding to the (bright) dipolar resonances of the two rods (see also the insets in Supplemental Figure S2 for $L_{2}=125\,\mu$m).  However, significant interference between them is observed in the dimer spectra, which do not correspond to simply a linear combination of the isolated spectra of each rod. Rather, rod interference leads to a stronger/weaker impact of the higher/lower frequency rod resonance (apart from their opposite phases, not shown), as expected \cite{Bozhevolnyi2011}. Still, to explain the two (bright and dark) array modes shown in Fig.~\ref{fig:DDexp1}, the impact of the lattice has to be  fully accounted for.

\textit{Theoretical model.}\textemdash To shed light onto the rich phenomenology that arises in the dimer arrays, we developed a simple coupled dipole-dimer model of an infinite array embedded in a homogeneous environment \cite{Abujetas2018a}. Dipoles are fully characterized by their polarizabilities along the $y$-axis, namely $\alpha_{y}^{(1)}$ and $\alpha_{y}^{(2)}$, where $(1)$ and $(2)$ account for each dimer dipole in the unit cell. The array is excited by an external plane wave, $\psi_{0}$ being its electric field polarization along the $y$-axis. Upon imposing  Bloch's theorem, the problem is reduced to finding the fields in one unit cell. The local field at the position of the dipoles, $\psi_{loc}^{(i)}$, with $i = 1,2$, can be found through a self-consistent field equation
\begin{equation}
\begin{bmatrix}
\psi_{loc}^{(1)} \\
\psi_{loc}^{(2)}
\end{bmatrix}
 = \left[\overleftrightarrow{I} - k^{2}\overleftrightarrow{G}_{b}\overleftrightarrow{\alpha}\right]^{-1} 
\begin{bmatrix}
\psi_{0}^{(1)} \\
\psi_{0}^{(2)}
\end{bmatrix},
\label{eq:sce}
\end{equation}
where $\psi_{0}^{(i)}$ is the incident field on dipole $i=1,2$, $\overleftrightarrow{I}$ is the identity matrix, $\overleftrightarrow{\alpha}$ is the polarizability tensor and $\overleftrightarrow{G}_{b} $ is the lattice ‘depolarization’ dyadic (or return Green function)
\begin{equation}
\overleftrightarrow{\alpha} = 
\begin{bmatrix}
\alpha_{y}^{(1)} & 0 \\
0 & \alpha_{y}^{(2)}
\end{bmatrix},
%\end{equation}
%\begin{equation}
\quad
\overleftrightarrow{G}_{b} = 
\begin{bmatrix}
G_{byy} & G_{yy}^{(1-2)} \\
G_{yy}^{(2-1)} & G_{byy}
\end{bmatrix}.
\end{equation}
$G_{byy}$ describes the self-interaction of each dipole array, i.e., the depolarization effect of (1) over (1), same as the effect of (2) over (2).  $G_{yy}^{(1-2)}$ and $G_{yy}^{(2-1)}$ are the interaction of the dipole array labeled as (1) over (2) and vice versa. Due to symmetry, at normal incidence $G_{yy}^{(1-2)} = G_{yy}^{(2-1)}$. 

Surface lattice resonances  are the solutions of Eq.~(\ref{eq:sce}) in the absence of the external plane wave.  To solve it, we diagonalize the system and find the condition at which the eigenvalues, $\Lambda$, are equal to zero. At real frequencies, the zeros of the real part of $\Lambda$ give the resonance frequencies, whereas the imaginary parts define the resonance widths. Recall that the imaginary components of $\overleftrightarrow{G}_{b}$ and $\overleftrightarrow{\alpha}$ (for lossless particles) are well defined and satisfy the following condition
\begin{equation}
\Im\left[\left(\dfrac{1}{\alpha_{y}} - G_{byy}\right) + G_{yy}^{(1-2)} \right] = 0,
\label{eq:imag}
\end{equation} 
where we define the magnitudes
\begin{equation}
\dfrac{2}{\alpha_{y}} = \dfrac{1}{k^{2}}\left(\dfrac{1}{\alpha_{y}^{(1)}} + \dfrac{1}{\alpha_{y}^{(2)}}\right),\quad \Delta\alpha_{y} = \dfrac{1}{k^{2}}\left(\dfrac{1}{\alpha_{y}^{(1)}} - \dfrac{1}{\alpha_{y}^{(2)}}\right) ,
\end{equation}
$\Delta\alpha_{y}$ thus representing the detuning between the two rods in the unit cell (introduced in the experiments by changing the size of one rod). For a small detuning, $\Delta\alpha_{y} \ll 2G_{yy}^{(1-2)}$, the imaginary components of the eigenvalues can be approximated by 
\begin{subequations}
\begin{eqnarray}
&&\Im\left[ \Lambda^{+} \right] = \Im\left[ \dfrac{\left(\Delta\alpha_{y}\right)^{2}}{8G_{yy}^{(1-2)}} \right],
\label{eq:lm} 
\\
&& \Im\left[ \Lambda^{-} \right] = \Im\left[ 2\left( \dfrac{1}{\alpha_{y}} - G_{byy} \right) - \dfrac{\left(\Delta\alpha_{y}\right)^{2}}{8G_{yy}^{(1-2)}}\right]. 
\label{eq:lma}
\end{eqnarray}
\end{subequations}
Finally, the corresponding eigenvectors are given by
\begin{equation}
\Lambda^{\pm} = 0 \rightarrow \nu^{\pm} = 
\begin{bmatrix}
\psi_{loc}^{(1)} \\
\psi_{loc}^{(2)}
\end{bmatrix} =
\begin{bmatrix}
1 \\
\mp 1
\end{bmatrix}
\end{equation} 
The lattice resonances are associated to two modes in which the rods are out-of-/in-phase (antisymmetric/symmetric). 
From Eq.~(\ref{eq:lm}), it follows that the imaginary part of $\Lambda^{+}$ goes to zero as the detuning is suppressed, i.e. as the two rods are made equal. Hence, as expected, the out of phase mode $\Lambda^{+}$ is very narrow and becomes a BIC at zero detuning. On the other hand, $\Im\left[ \Lambda^{-} \right]$ is always larger than zero, so that the in-phase mode is broad. Therefore, for $L_{1} \not = L_{2}$ ($\alpha_{y}^{(1)} \not = \alpha_{y}^{(2)}$) we have a broad mode that interferes with a very narrow mode, leading to a Fano resonance. For $L_{1} = L_{2}$, the narrow mode converges into a BIC state, precluding any external coupling to it and the formation of the dip in transmission.   

Although we have considered that the rods are separated along the $x$-axis, if there is also an additional displacement along the $y$-axis, given by $d_y$, Eq.~(\ref{eq:imag}) still holds. Moreover, this identity is also valid for non-square lattices where $a \not = b$. Hence, Eq.~(\ref{eq:imag}) is universal, as long as there is no coupling involving higher multipoles, and holds for: (i) any set of lattice constant parameters, $a$ and $b$; and (ii) any relative displacement between dipoles inside the unit cell, not only along the $x$-axis, but extensive to the full $xy$ plane. \textit{Therefore, a major conclusion is that the BIC state is symmetry-protected and robust against changes in the specific lattice parameters: $a$, $b$, $d_x$ and $d_y$. }
\begin{figure}
\includegraphics[width=1\columnwidth]{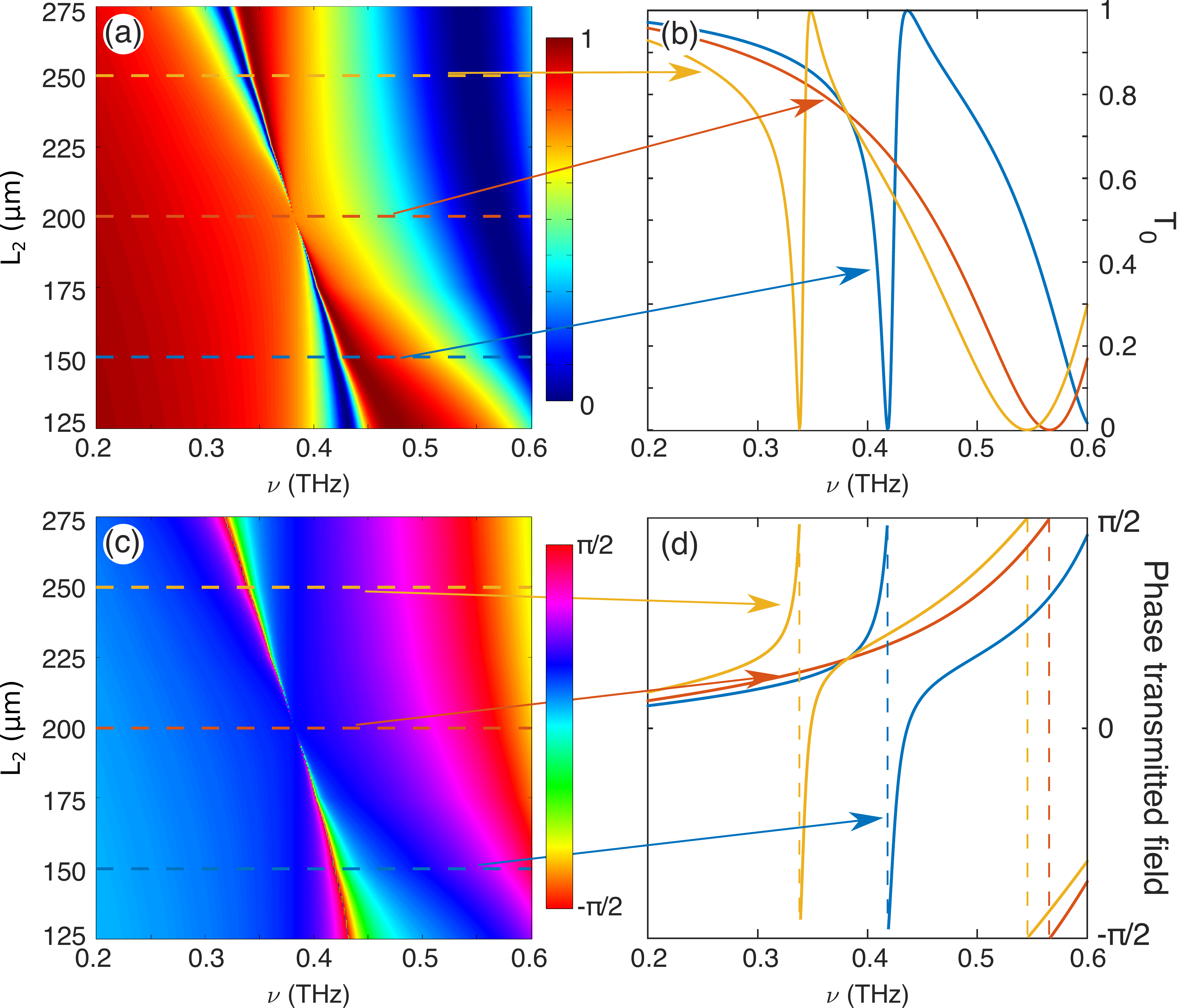}
\caption{(a-d) Theoretical transmittance spectra calculated through coupled dipole theory for a square lattice ($a=300$ $\mu$m) as in Fig.~\protect{\ref{fig:DDexp1}}, but consisting of two detuned dipoles per unit cell, separated by $d_x = 2a/5$: (a,b) Transmittance; (c,d) phase.  Dipole polarizibilities are extracted from the numerically calculated scattering cross sections shown in Supplemental Fig.~S2 (see text):  One is fixed and corresponds to a rod with dimensions  L$_1=$200 $\mu$m and $w_1=$40 $\mu$m, while the other one $L_{2}$ (with identical areas $L_2w_2=L_1w_1$) varies continuously in the contour maps in (a,b), whereas those cases corresponding to three of the experimental dimers,  $L_{2}(\mu$m)$ =150, 200, 250$, are shown in (b,d).}
  \label{fig:DDtheo}
\end{figure}

Let us now analyze the Fano-BIC transition using our coupled dipole-dimer model. We plot in Fig.~\ref{fig:DDtheo}(a,c) the spectra of the transmission coefficient ($T=1-R_{0}$) intensity and phase, for a square lattice (lattice constant $a=b=300$ $\mu$m) with two dipoles per unit cell, separated by a distance of d$_x = 120$ $\mu$m for varying $L_2$; cuts for fixed lengths are shown in Fig.~\ref{fig:DDtheo}(b,d). 
The polarizability of the rods is calculated through  SCUFF \cite{SCUFF1,SCUFF2}, considering the rods as perfect electric conductors [see Supplemental Fig.~S2(b)]. In order to avoid numerical problems related to unphysical absorption, we consider  lossless dipoles  and fix the imaginary parts of the polarizabilities to $\Im\left[1/\alpha_{y}^{(i)} \right]= -k^{3}/(6\pi) $ to fulfill the optical theorem, taking for the real parts the values calculated numerically. 
%Above $\nu = 0.64$ THz, the first diffraction channel is open for the analytical model and cannot be directly compared to the experimental results. 

Figure~\ref{fig:DDtheo} shows that the coupled dipole-dimer model reproduces all the features exhibited in Fig.~\ref{fig:DDexp1}, fully extending the characterization of the BIC into the $L_2$-parameter space. Remarkably, the contour map in Fig.~\ref{fig:DDtheo}(a) reveals the classical narrowing of the leaky resonance (Q-factor tending to infinity) towards a BIC state, with the distinct feature that the resonant state [surface lattice dipole resonance with an abrupt $\pi$-phase jump shown in the contour map in Fig.~\ref{fig:DDtheo}(c)] manifests itself as a narrow Fano resonance instead.  Transmittance spectra (intensity and phases) are shown in Fig.~\ref{fig:DDtheo}(c,d) for given $L_2\not = L_1$  illustrating this Fano-like behavior that disappears at    $L_2= L_1$  as a signature of the BIC.

\begin{figure}
\includegraphics[width=1\columnwidth]{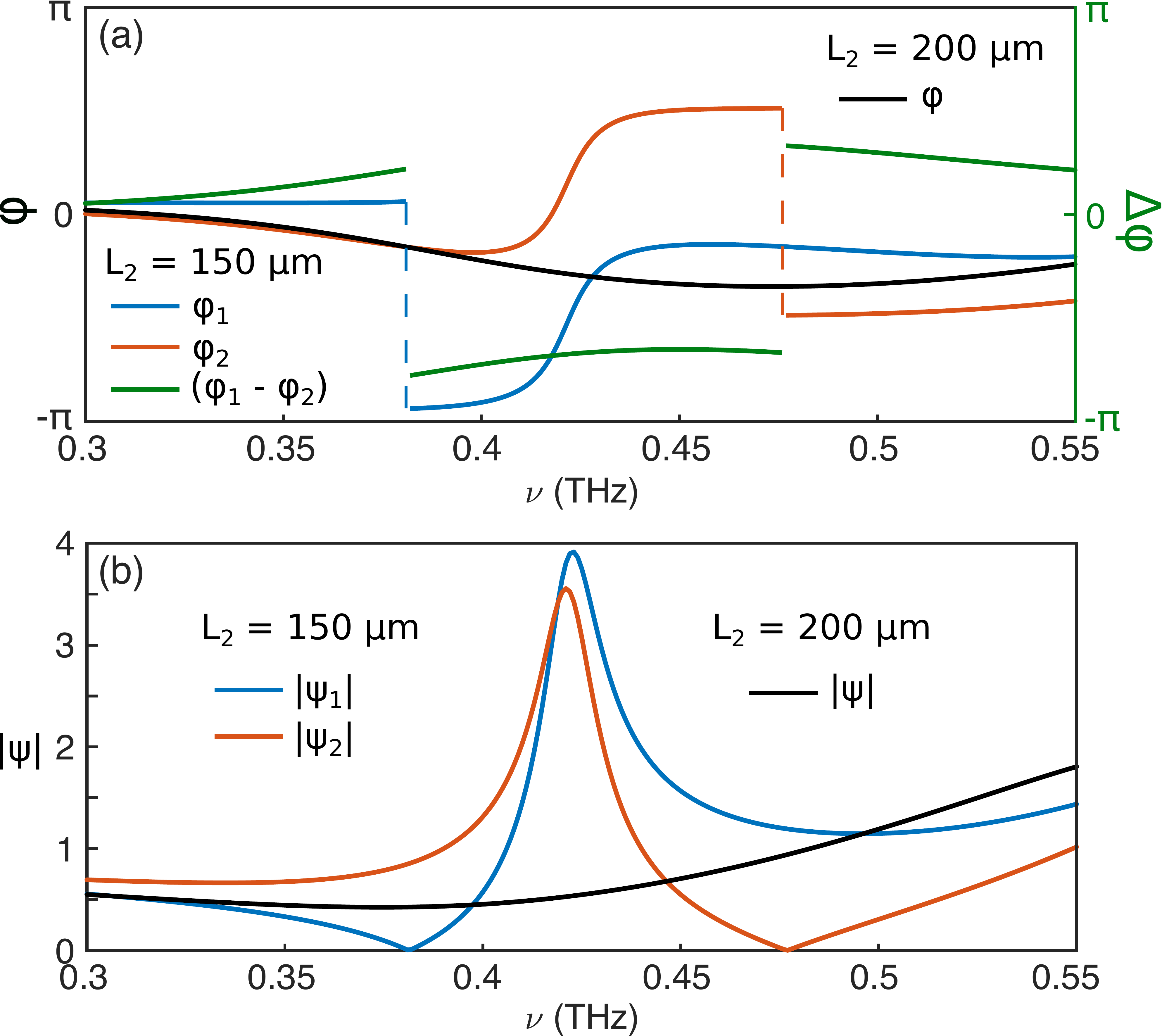}
\caption{Spectral variation of both local field (a) phases (including the relative phase)  and (b) amplitudes, calculated through coupled dipole theory for a square lattice ($a=300$ $\mu$m) as in Fig.~\protect{\ref{fig:DDexp1}}, but consisting of two detuned dipoles per unit cell, separated by $d_x = 2a/5$ for $L_{2}=$200 $\mu$m and $L_{2}=$150 $\mu$m. The local amplitudes and phases at both rods for $L_{2}=200 $ $\mu$m are identical.}
  \label{fig:DDtheoloc}
\end{figure}

To shed more light onto the phenomenology, we show in Fig.~\ref{fig:DDtheoloc} the amplitudes ($\psi_{loc}^{(1)},\psi_{loc}^{(2)}$) and phases ($\phi_{loc}^{(1)},\phi_{loc}^{(2)},\Delta\phi=\phi_{loc}^{(1)}-\phi_{loc}^{(2)}$) of the local fields over the dipoles,  for the dipole dimer corresponding to  $L_{2} = 150$ $\mu$m.  At low frequencies, both dipoles are driven in phase. At $\nu = 0.38$ THz, coinciding with the zero of the field of one dipole, $\Delta\psi$ presents a discontinuity and the phase difference becomes maximum (rods out of phase, with  a  high dispersion in both). The field amplitudes are enhanced and, indeed, exhibit a resonant lineshape, corresponding to the dark lattice resonance, where the Fano asymmetric lineshape emerges in the far field. Then at $\nu = 0.48$ THz, upper-frequency end of the Fano resonance, $\Delta\psi$ presents another discontinuity exactly at the zero of the field amplitude of the other dipole, see again Figs.~\ref{fig:DDtheoloc}; for higher frequencies the phase difference vanishes and the dipoles are in phase again. Thus, at given frequencies at the lower/upper band of the Fano resonance, the fields at the long/short rods are strictly zero, manifesting the strong interaction that exist between the different rods near resonance. In addition, the local field is enhanced by more than a factor of three in both rods. For values of $L_{2}$ closer to $L_{1}$, these features become more prominent. At $L_{2} = L_{1}$, both dipoles are equal and the anti-phase behavior is not allowed, so that the field amplitude/phase at both rods is identical [cf. Figs.~\ref{fig:DDtheoloc}(a,b)], with no evidence whatsoever of a Fano resonance, thus becoming a BIC [as shown in Fig.~\ref{fig:DDtheo}(a,b)].

\begin{figure}
\includegraphics[width=1\columnwidth]{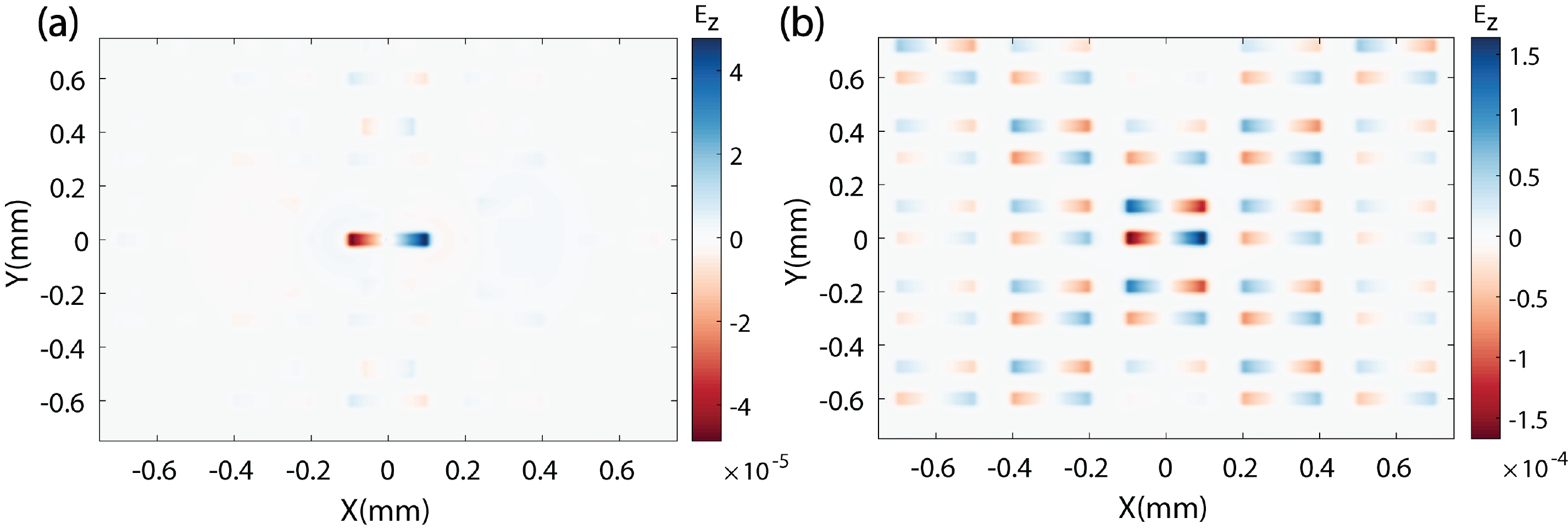}
\caption{Near Field map showing the electric field component along the $z$-direction, numerically calculated through Lumerical, for a dipole source (located at the image center) inside a square lattice  ($a=300$ $\mu$m) of  Au rod dimers deposited on a quartz substrate, with  rod separation $d_x = 2a/5$. One of the rods has  fixed dimensions of  L$_1=200\,\mu$m and $w_1=40\,\mu$m, while the other is (a)  $L_{2}=125\;\mu$m (i.e. detuned dipoles) and (b) $L_{2}=200\;\mu$m (i.e. equal dipoles), while keeping the surface area fixed: $L_2w_2=L_1w_1$. The dipole source emits near the BIC frequency at $\nu=0.357$ THz.}
  \label{fig:FDTDBIC}
\end{figure}

To reveal even more neatly the BIC behavior, we show a near-field map in Fig.~\ref{fig:FDTDBIC}, numerically calculated through FDTD (Lumerical), for a dipole source (located at the image center) inside a dimer rod array ($d_x=120$ $\mu$m) with identical dimensions, $L_{2}=L_{1}=200\;\mu$m and $w_1=w_2=40\;\mu$m, at the BIC frequency $\nu=0.357$ THz [Fig.~\ref{fig:FDTDBIC}(b)]; another case with different rods $L_{2}=125\;\mu$m is also shown for comparison [Fig.~\ref{fig:FDTDBIC}(a)]. In Fig.~\ref{fig:FDTDBIC}(a) the excitation with a point dipole is radiated to the far field almost immediately, the near field map revealing that only the rod dimer adjacent to it is excited;  in particular, the resonant rod with $L_{1}=200\;\mu$m, as expected. By contrast, the near field for equal rod dimers is effectively trapped in the BIC mode as shown in Fig.~\ref{fig:FDTDBIC}(b): many dimers in the lattice are resonantly excited, with the expected opposite phase for each rod within the (dimer) unit cell that precludes out-of-plane radiation losses. 

\textit{Conclusions.}\textemdash We have shown experimentally through the THz transmission spectra  that metasurfaces consisting of asymmetric, sub-wavelength gold rods exhibit strong Fano resonances, which become narrower as the dimensions of the rods approach each other, turning into BICs when they are identical. Such experimental results have been fully explained in the theoretical context of detuned-dipole arrays. This evolution manifests through a spectral map in the (detuning) parameter space with asymmetric Fano lineshapes tending to an infinite Q-factor as the condition for the BIC emergence is approached. In addition, it is explicitly shown that the condition to support a BIC is independent  of both the relative position between  rods inside the unit cells and  the lattice constants (provided that no diffractive orders come into play).  Hence, these properties make an array of detuned dipoles a general scenario to engineer robust and versatile metasurfaces supporting Fano resonances and bound states in the continuum throughout the electromagnetic spectrum, with appealing implications in sensing, lasing, and related phenomenology. Finally, bear in mind that the formalism and phenomenology of our coupled detuned-dipole model could be extrapolated to other fields of wave physics with arrays of dipolar scatterers, such as acoustic, elastic, seismic and even atom waves.

\acknowledgments
We thank M. Ramezani for fruitful discussions. This work has been supported in part by the Spanish Ministerio de Ciencia, Innovaci\'on y Universidades (LENSBEAM FIS2015-69295-C3-2-P, NANOTOPO FIS2017-91413-EXP, and FPU PhD Fellowship FPU15/03566); and by the Nederlandse Organisatie voor Wetenschappelijk Onderzoek (NWO) through the Industrial Partnership Program NanoPhotonics for Solid State Lighting and the Innovational Research Incentives Scheme (NWO-Vici grant SCOPE nr. 680-47-628)

%\bibliography{library}


\end{document}